# Charging effects in biased molecular devices

Kamil Walczak [1]

Institute of Physics, Adam Mickiewicz University
Umultowska 85, 61-614 Poznań, Poland

The influence of the charging effects on the transport characteristics of a molecular wire bridging two metallic electrodes in the limit of weak contacts is studied by generalized Breit-Wigner formula. Molecule is modeled as a quantum dot with discrete energy levels, while the coupling to the electrodes is treated within a broad band theory. Owing to this model we find self-consistent occupation of particular energy levels and orbital energies of the wire in the presence of transport. The nonlinear conductance and current-voltage characteristics are investigated as a function of bias voltage in the case of symmetric and asymmetric coupling to the electrodes. It is shown that the shape of that curves are determined by the combined effect of the electronic structure of the molecule and by electron-electron repulsion.

## I. Introduction

In the last decade, magnificent progress has been made in the theoretical and experimental investigations of molecular devices [1-3]. Molecular junctions have attracted considerable interest due to their small sizes, self-assembly features, mechanical flexibility, chemical tunability and unique transport properties. Until now, a lot of such structures have been proposed as a useful components of electronic circuits and realized in practice. The junction is usually made of two planar metallic electrodes joined by a molecular wire. The charge is transferred through the molecular bridge under the influence of bias voltage and current-voltage (I-V) dependences are measured. Transport properties of the junction strongly depend on the quantum nature of molecular system, electronic structures of the electrodes near Fermi level and the molecule-to-electrodes coupling [4]. Additionally, it is expected that the electron-electron interaction becomes important at the atomic scale and may produce essential effects in the transport characteristics [5]. In fact, negative differential resistance (NDR) has been reported [6] and the suggested theoretical explanation is based on charging of the molecule [7].

The main purpose of this work is to study changes of the transport characteristics (conductance spectra and current-voltage dependences) due to the charging effects. A convenient framework for analyzing this problem in the self-consistent manner is based on the generalized Breit-Wigner formula [8-14]. In order to reproduce the experimental values of the current we must postulate that the molecular wire is weakly coupled to the electrodes [15]. However, among all the factors that significantly affect the strength of the coupling we can enumerate: the geometry of the contacts, the type of molecule-to-electrodes binding and the molecule-electrode distance [12,16]. Since the effective coupling is weak, molecule is modeled as a quantum dot with discrete energy levels, while both electrodes are treated as semi-infinite reservoirs of free electrons at thermal equilibrium (see Fig.1).

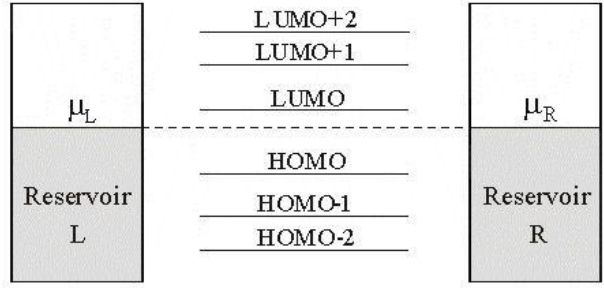

Fig.1 Energy level diagram for a molecular device.

In this simplified picture, the current is due to sequential electron transfer, where the molecular wire is successively charge and discharged by the tunneling process. Although presented model does not allow us to calculate electrostatic potential profile along the wire, it provides desired information concerning the charging effects. The charging of the molecule at large bias greatly influences the behavior of the conductance, as will be discussed later in this paper.

## II. Description of the model

As mentioned above, the device is made of two electrodes joined by a molecular wire. If we restrict our considerations to weak-coupling limit, particular energy levels of the wire are well-separated (i.e. do not overlap with each others) and therefore the transmission can be written as a summation over all the states (for two spin directions) [12-14]:

$$T(\omega) = \sum_r \frac{4\Delta_{rL}\Delta_{rR}}{[\omega - \varepsilon_r]^2 + [\Delta_{rL} + \Delta_{rR}]^2}, \quad (1)$$

where: $\varepsilon_r$ enumerates the energies of the effective molecular orbitals contributing to transport. Particular energy levels are redefined by incorporating "Hartree contributions" to the "bare" molecular levels [17]:

$$\varepsilon_r = \varepsilon_r^{(0)} + U \sum_\sigma <n_{r,\sigma}>, \quad (2)$$

where $\varepsilon_r^{(0)}$ denote "bare" energies. Here we have no reason to distinguish spin polarization and therefore we assume that both configurations contribute in equal part to the state occupation. However, spin-dependent transport occurs in the presence of ferromagnetic electrodes [18,19].

In our case, Coulomb interaction is described via an on-level repulsion of strength U and the thermal average for occupation numbers $<n_{r,\sigma}>$ can be computed using Keldysh formalism [20,21]:

$$<n_r> = \frac{1}{\pi} \int_{-\infty}^{+\infty} d\omega \frac{\Delta_{rL} f(\omega - \mu_L) + \Delta_{rR} f(\omega - \mu_R)}{[\omega - \varepsilon_r]^2 + [\Delta_{rL} + \Delta_{rR}]^2}, \quad (3)$$

where: $\Delta_{rL/rR}$ is the coupling parameter of r-orbital to L/R reservoir, $f(\omega - \mu_{L/R})$ denotes Fermi distribution function with electrochemical potentials $\mu_{L/R} = \varepsilon_F \pm eV/2$. In practice, spin degrees of freedom are simply taken into account by doubling electron population for two possible spin configurations in eq.3.

Since the molecular energy level $\varepsilon_r$ depends on the occupation number and the occupation number is defined through the energy $\varepsilon_r$, we need to recalculate such quantity in the self-consistent procedure until convergence. First we intelligently choose occupation numbers of particular levels (between 0 and 2) and then calculate their energies (eq.2). Having energies we can recalculate the population of charge carriers in our computational scheme (eq.3). That procedure is continued until the time when the next step does not change all the energies of molecular levels significantly. Such energies are later used to receive the current for given bias voltage (eq.4). The whole procedure is repeated for different values of applied bias.

The expectation value of the current flowing through the device is related to the transmission function [20,21]:

$$I(V) = \frac{2e}{h} \int_{-\infty}^{+\infty} d\omega T(\omega)[f(\omega - \mu_L) + f(\omega - \mu_R)]. \qquad (4)$$

The differential conductance is then obtained from the current formula as its derivative with respect to voltage. In calculations, we have assumed that transport is elastic and electron-electron interaction does not destroy coherence.

**III. Results and their interpretation**

Different types of nanowires are possible to provide, although the most popular are organic molecules [1], where conduction is due to π-conjugated molecular orbitals. Such orbitals extend over the whole molecule and facilitate the transport of electrons between two reservoirs. In particular, linear carbon-atom chains containing up to 20 atoms connected at the ends to metal atoms have been synthesized [22] and recognized as ideal one-dimensional wires [23,24].

As an example, we have studied linear chain containing 12 carbon atoms weakly connected at the ends to the metal electrodes [25]. Since only π-electrons are involved into the conduction problem, initial electronic structure of the molecule is obtained through the use of Hückel Hamiltonian (with orthogonal atomic basis set of states) in the procedure of diagonalization [26] ("bare" energy levels are given in eV: $\pm 0.603$, $\pm 1.773$, $\pm 2.840$, $\pm 3.743$, $\pm 4.427$, $\pm 4.855$). In the strict Hückel approach, conducting energy level of the molecule produces a step-like contribution to the current. Therefore, the I-V function should look as a series of steps, occurring when the resonant conditions are satisfied for particular conducting orbitals. For further calculation we assume realistically that Fermi level of unbiased electrodes is closer to the LUMO level ($\varepsilon_F = 0.3$) [15,27]. Temperature of the system is set to 293 K, but anyway our results are not particularly sensitive to temperature.

Figure 2 presents transport characteristics for 12-atom wire, when the connection to the electrodes is the same at both ends. In the symmetric-coupling case, there is no physical reason to appear any asymmetry in the spectra. Indeed, both curves (current-voltage and conductance-voltage) reveal symmetric behavior independently from the strength of the electron-electron repulsion. Besides, the inclusion of charging results in modification of the steps in the I-V dependence, which gain finite slope. Generally, the slope of the step decreases with the increase of U-parameter and I-V curve becomes fairly smooth.

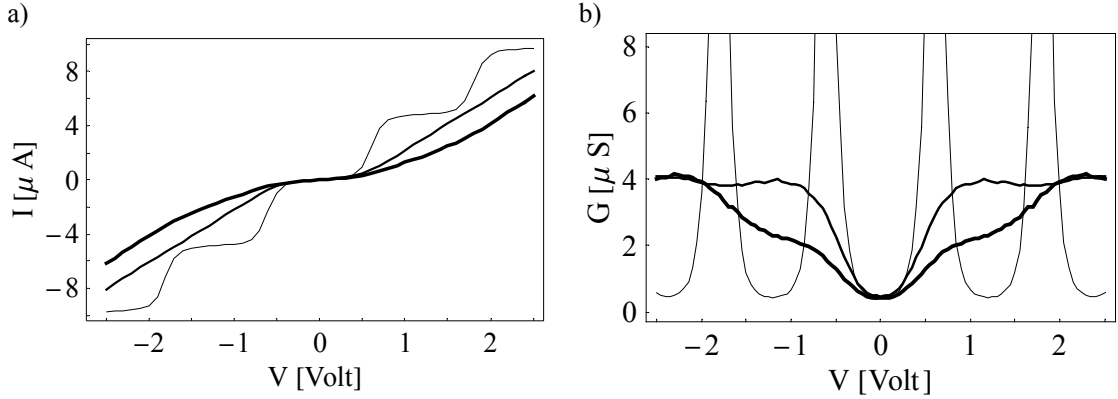

Fig.2 Results for linear chain of 12 carbon atoms for symmetric coupling to the electrodes ($\Delta_L = 0.01 = \Delta_R$) and different values of U = 0,1,2 (with the increase of linewidth): (a) current-voltage and (b) conductance-voltage characteristics.

For sufficiently large U, the current steps are joint together into one continuous line and I-V characteristic does not display its jump-like character (see Fig.2a).

On the other hand, charging enters the picture of conductance only at higher voltages, when an electrochemical potential tries to cross the conducting level, completely changing its behavior. The inclusion of electron-electron repulsion tends to broaden the sharp peaks of conductance, even if we have not included any extra level broadening into our computational scheme (such as: thermal or additional coupling effects). For larger U, peaks in conductance spectrum are shifted in the direction of higher voltages and their height is lowered. In addition, these peaks gain substantial width and may even merge (see Fig.2b).

In Fig.3 we plot transport characteristics for 12-atom wire, where the connection to the electrodes is different at both ends. Our predictions indicate that there is no possibility to generate asymmetric transport dependence in the case of negligence of Coulomb interactions (although the molecule-to-electrodes coupling is asymmetric). Here we observe the effect of U-induced current rectification, i.e. the magnitude of the current depends on the polarity of bias voltage. Moreover, asymmetry in the I-V behavior increases with increasing U-parameter (see Fig.3a). Such diode-like characteristic has been observed in a number of experiments involving monolayer and multilayer films as well as a single molecule type junctions and STM measurements [28,29]. Within our model, the rectification effect is explained as a combined effect of asymmetry coupling to the reservoirs and electron-electron repulsion. U-induced asymmetry is also seen in the conductance spectra (see Fig.3b).

**IV Concluding remarks**

In this paper, we have proposed a simple model to address the question of charging effects in biased molecular devices, when the molecule is weakly coupled to the electrodes. Our model is based on the Breit-Wigner type formula for the resonant conductance. The calculations predict significant changes in transport characteristics due to the inclusion of Coulomb interactions in a self-consistent way. It is shown that the shape of current-voltage and conductance-voltage curves are determined by the combined effect of the electronic structure of the molecule and by electron-electron repulsion. In particular, the rectification effect stems from simultaneous considerations of asymmetry coupling to the reservoirs and Coulomb interactions.

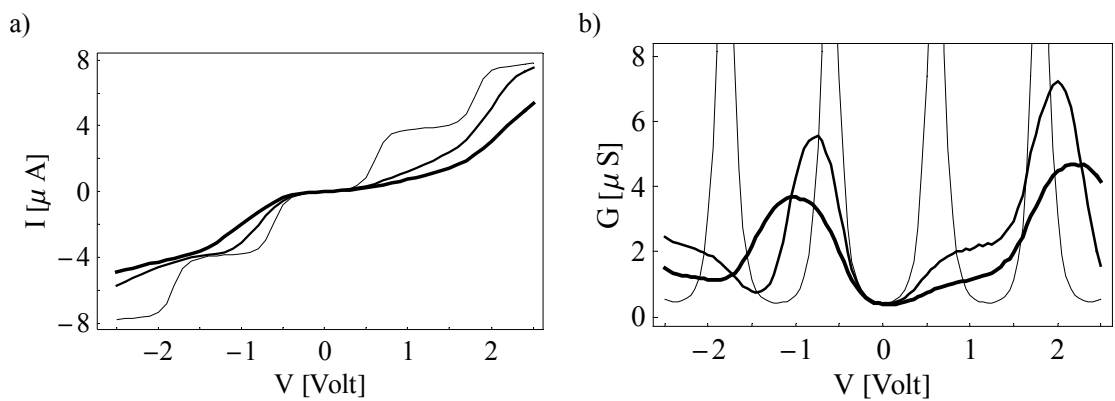

Fig.3  Results for linear chain of 12 carbon atoms for symmetric coupling to the electrodes ($\Delta_L = 0.02 = 4\Delta_R$) and different values of $U = 0,1,2$ (with the increase of linewidth): (a) current-voltage and (b) conductance-voltage characteristics.

It should be also noted that final conclusions are in agreement with some recent investigations [30]. Moreover, charging-induced asymmetry of the I-V curve was suggested earlier [31], but in a slightly different approach to the problem of electron conduction [32].

In conclusion, it is important to stress some limitations of our model. Presented model fails when the width of the molecular orbitals is comparable with energy difference between them, so that the resonances may overlap (the strong-coupling limit). Furthermore, transport is analyzed as a coherent tunneling, where the transit time is much less than the intramolecular vibrations. However, for organic molecules that condition could not be satisfied [33]. In this situation relaxation processes become important and should be taken into account. Anyway, dephasing effects are far beyond the scope of the present work.

**Acknowledgments**

The author is very grateful to B. Bułka, T. Kostyrko and B. Radzimirski for many interesting discussions. This work was supported in part by the State Committee for Scientific Research (Poland) within the project No. PBZ KBN 044 P03 2001 and in part by the European Commission under the contract No. MRTN-CT-2003-504574.